\documentclass{ws-procs9x6}

\usepackage[utf8]{inputenc}
\usepackage[T1]{fontenc}

\begin{document}

\title{SPIN DEPENDENT TRANSPORT IN AN ELECTRON GAS WITH MAGNETIC DISORDER}

\author{T. L. van den Berg$^*$ and A. D. Verga}

\address{IM2NP, Aix-Marseille Université,\\
Marseille, 13001, France\\
$^*$E-mail: tineke.vandenberg@im2np.fr\\
www.im2np.fr}

\begin{abstract}
We consider a two dimensional semiconductor with carriers subject to spin-orbit interactions and scattered by randomly distributed magnetic impurities. We solve the time-dependent Schrödinger equation to investigate the relationship between the geometrical properties of the wavefunction and the system's spin dependent transport properties. Even in the absence of localized states, interference effects modify the carrier diffusion, as revealed by the appearance of power laws dependent on the impurity concentration. For stronger disorder, we find a localization transition characterized by a fractal wavefunction and enhanced spin transport.
\end{abstract}


\bodymatter

\section{Introduction}\label{s:intro}
An electron gas confined in a semiconductor heterostructure can display, under the action of an external electric field, spin dependent transport driven by the spin-orbit interaction.\cite{Sinova-2004fb} Although this intrinsic spin Hall effect is suppressed by disorder,\cite{Inoue-2004fk} magnetic impurities can restore it.\cite{van-den-Berg-2011fk}

Indeed, in the absence of disorder, an electric field in the \(x\)-direction, induces an out of plane spin polarization following the \(y\)-direction. The spin current \(j_y^z\) is related to the applied electric field \(E_x\) by the spin conductivity \(j_y^z=\sigma_{sH}E_x\), whose value \(\sigma_{sH}=-e/8\pi\) (\(e\) is the electric charge), turns out to be independent of the spin-orbit coupling constant \(\lambda\).\cite{Sinova-2004fb} Non-magnetic disorder preserves the time reversal symmetry of the microscopic Hamiltonian, and allows the setting of a macroscopic stationary state. Due to the particular form of the Rashba term, linear in the momentum, the spin current is proportional to the time derivative of the in plane spin; as this derivative must vanish at equilibrium, the spin Hall effect is canceled.\cite{Rashba-2004fk} In the presence of magnetic disorder, the system looses its microscopic time reversal invariance, allowing the emergence of the intrinsic spin Hall effect, even if this invariance is statistically restored.\cite{van-den-Berg-2011fk,Gorini-2008lq} 

Our purpose here is to investigate the behavior of the spinor wavefunction when the impurity concentration increases and to relate its dynamical and geometrical properties to the transport properties of the system. We present the model in Sec.~\ref{s:model}, followed be an account of the numerical results (Sec.~\ref{s:results}) and the conclusions (Sec.~\ref{s:concl}).

\section{Model}\label{s:model}

The Hamiltonian of the two-dimensional system writes
\begin{equation}\label{e:H}
H=\frac{p^2}{2m}-\frac{\lambda}{\hbar}\bm{\sigma}\cdot(\hat{\bm{z}}\times\bm{p})-J_s\bm{\sigma} \cdot \bm{n}(\bm{x})\,,
\end{equation}
where the second term represents the Rashba spin-orbit coupling, with parameter \(\lambda\), and the last term describes the exchange interaction, with exchange energy \(J_s\) and magnetic impurities \(i\),
\begin{equation}\label{e:n}
\bm n(\bm x) = a^2 \sum_{i\in \mathcal{I}} \bm n_i\delta(\bm x-\bm x_i)\,,
\end{equation}
oriented in the direction \(\bm n_i\), and distributed randomly; here \(\mathcal{I}\) is the set of impurity sites in the plane \(\bm x=(x,y)\) (\(a\) is a length constant); \(\bm \sigma\) is the vector of Pauli matrices, \(m\) the carrier's effective mass, and \(\bm p=(p_x,p_y)\) their momentum. We consider the magnetic impurities in a paramagnetic state, therefore the average of the last term in (\ref{e:H}), over the magnetic moment orientations \(\bm n_i\), vanishes.

The dynamics of the quantum state \(|\psi\rangle\) is governed by the Schrödinger equation,
\begin{equation}\label{e:Sch}
\mathrm{i}\frac{\partial}{\partial t}|\psi(t)\rangle=H|\psi(t)\rangle,
\end{equation}
where we chose units such that \(\hbar=a=m=1\). We numerically integrate (\ref{e:Sch}) using a Chebyshev spectral method,\cite{Weisse-2008uq} to obtain the time evolution of an initial spin-up gaussian wave packet. We measure the width,
\begin{equation}\label{e:w}
w^2_{\!\pm}(t)=\int_{L^2}\!d\bm x \,|\bm x|^2 |\psi_{\!\pm}(\bm x,t)|^2\,,
\end{equation} 
where \(\psi(\bm x,t)=\langle \bm x|\psi(t)\rangle=(\psi_{\!+}, \psi_{\!-})^T\) is the spinor whose components are the spin up and down amplitudes, and \(L^2\) is the system's surface. The initial orientation of the spin is not relevant, scattering off impurities is statistically isotropic, so that asymptotically both spin states display the same behavior. The spin current in the (arbitrary) direction \(y\), is given by
\begin{equation}\label{e:jyz}
j^z_y(t)=\frac{1}{4}\langle \psi(t)|\{\sigma_z,v_y\}|\psi(t)\rangle = 
	\frac{1}{2}\mathrm{Im} \! \int\! dx
		 \! \left(\psi^*_{\!+}(t) \partial_y \psi_{\!+}(t) - 
	\psi^*_{\!-}(t) \partial_y \psi_{\!-}(t) \right)\,,
\end{equation}
using the spinor in the momentum representation; here \(v_y\) is the \(y\) component of the velocity operator, and the brackets \(\{\cdot,\cdot\}\) are for the anti-commutator. The physical dimension is \([j^z_y]=[\hbar p/mL^2]\), the same as a force: charge (coming from the spin conductivity), multiplied by an electric field (driving force). In the absence of an external electric field, the transverse spin current should vanish in a stationary state, however in the present initial value problem, the \(|\psi(0)\rangle\) is not an eigenstate of \(H\), leading to an unbalance of spin-up and spin-down probability currents, which results in a (at least transitory) net spin current.

%wave
\begin{figure}[t]
\centering
\includegraphics[width=0.24\textwidth]{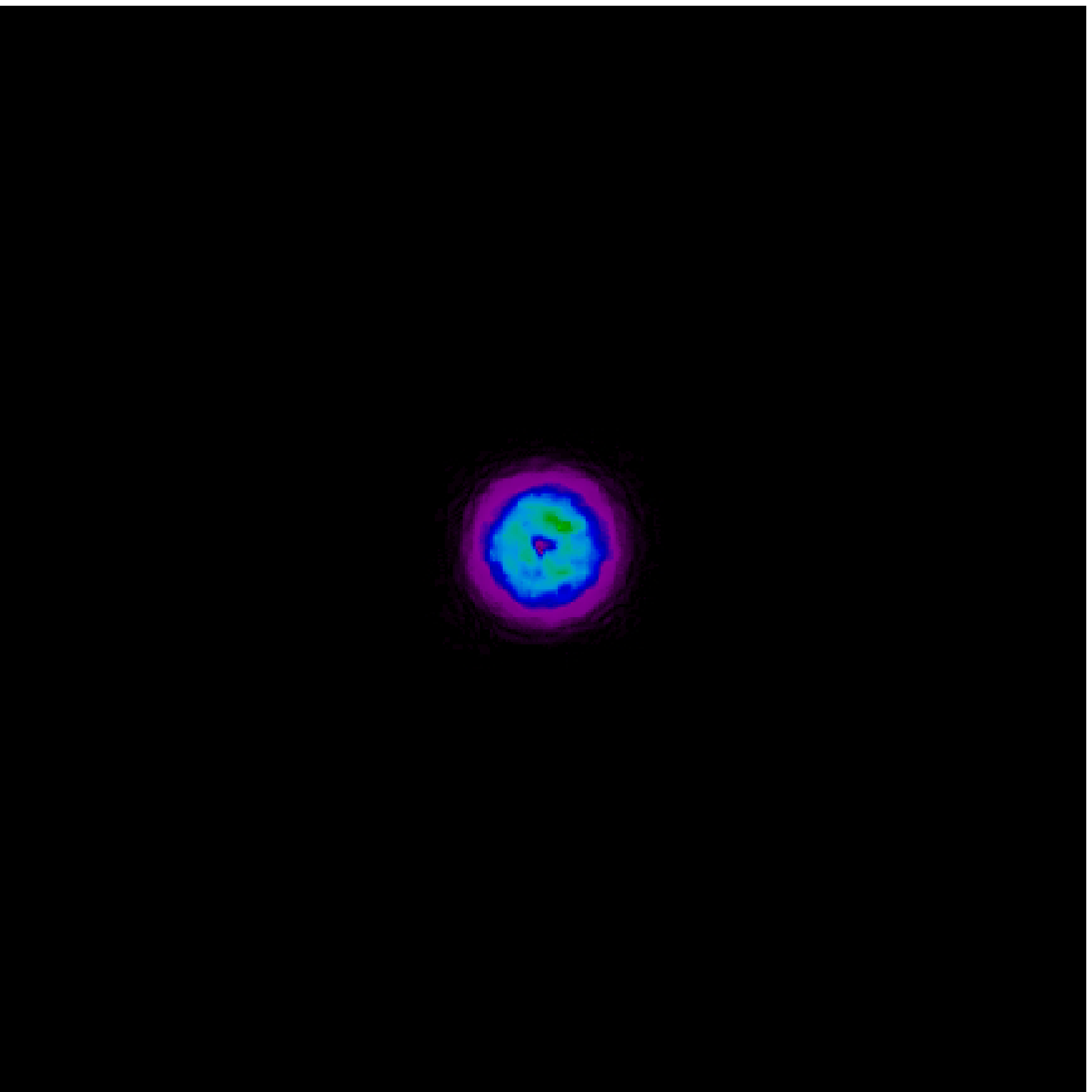}%
\includegraphics[width=0.24\textwidth]{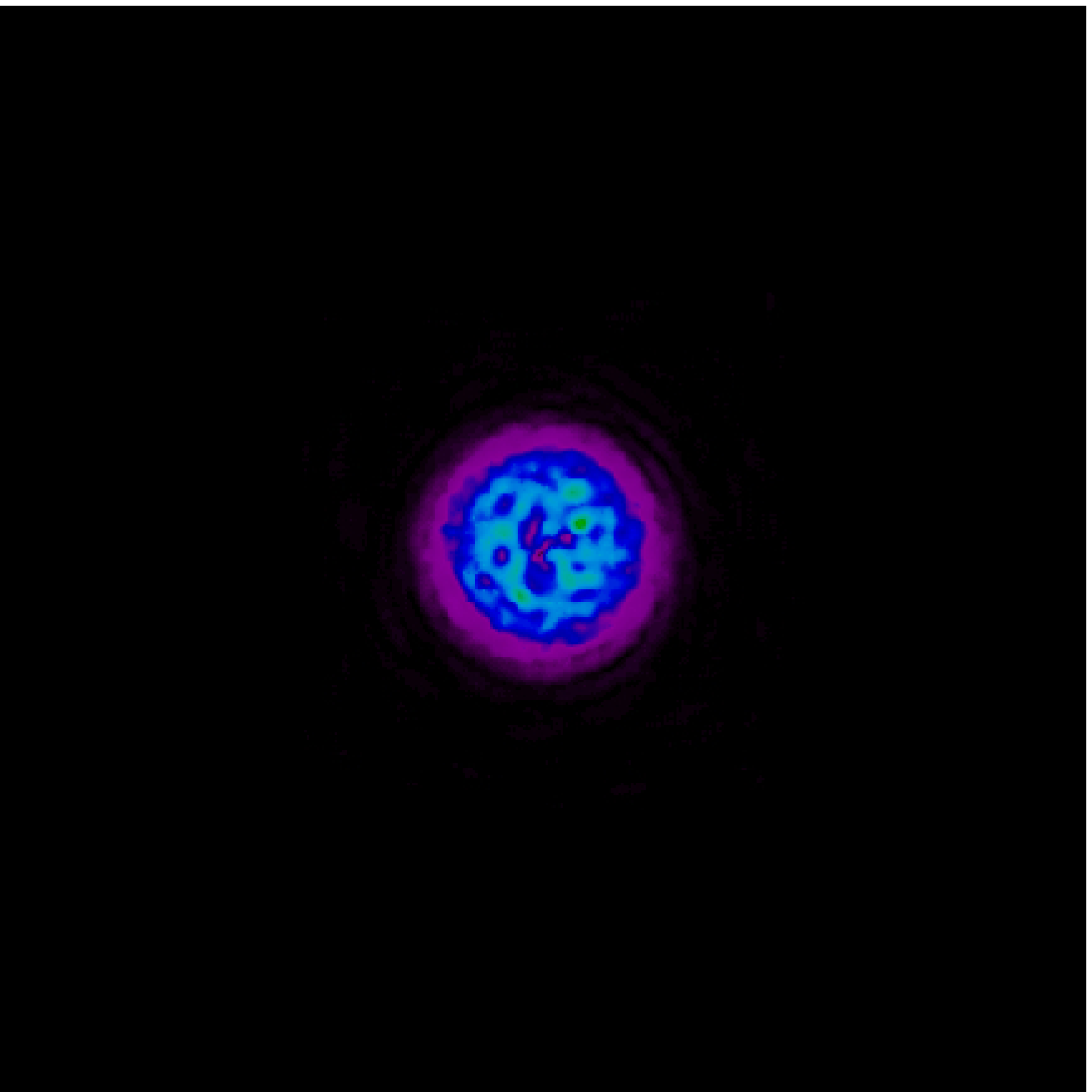}%
\includegraphics[width=0.24\textwidth]{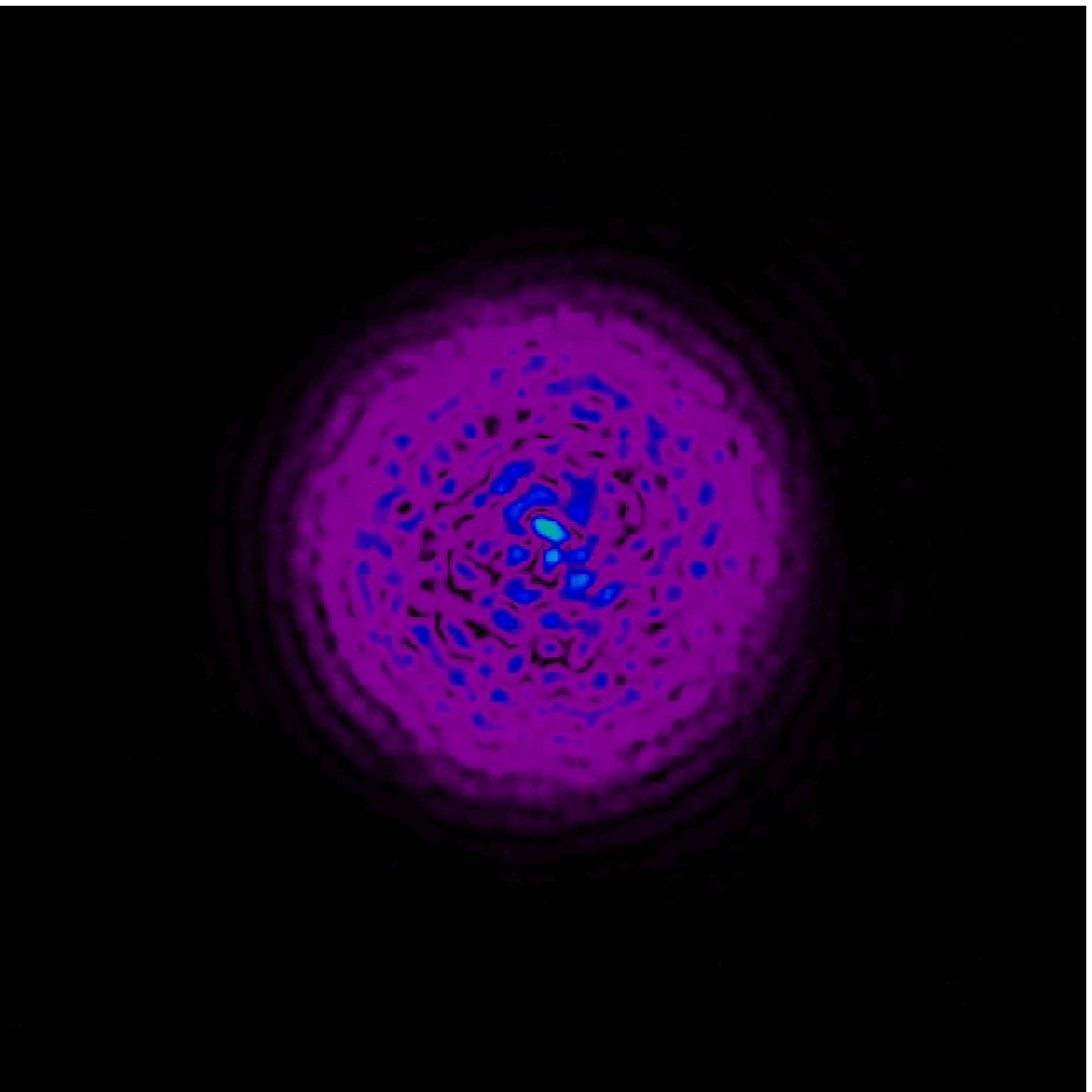}%
\includegraphics[width=0.24\textwidth]{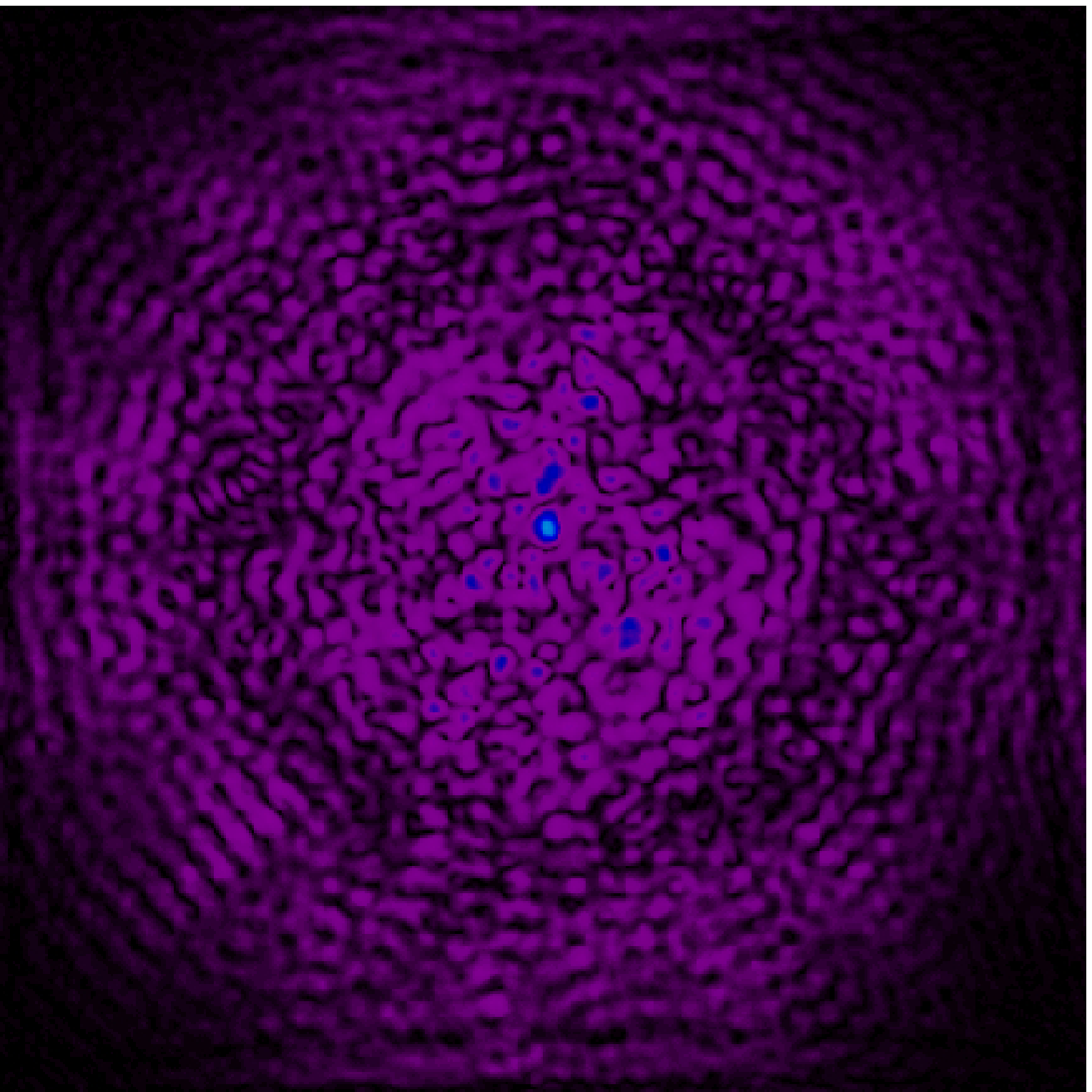}\\
\includegraphics[width=0.24\textwidth]{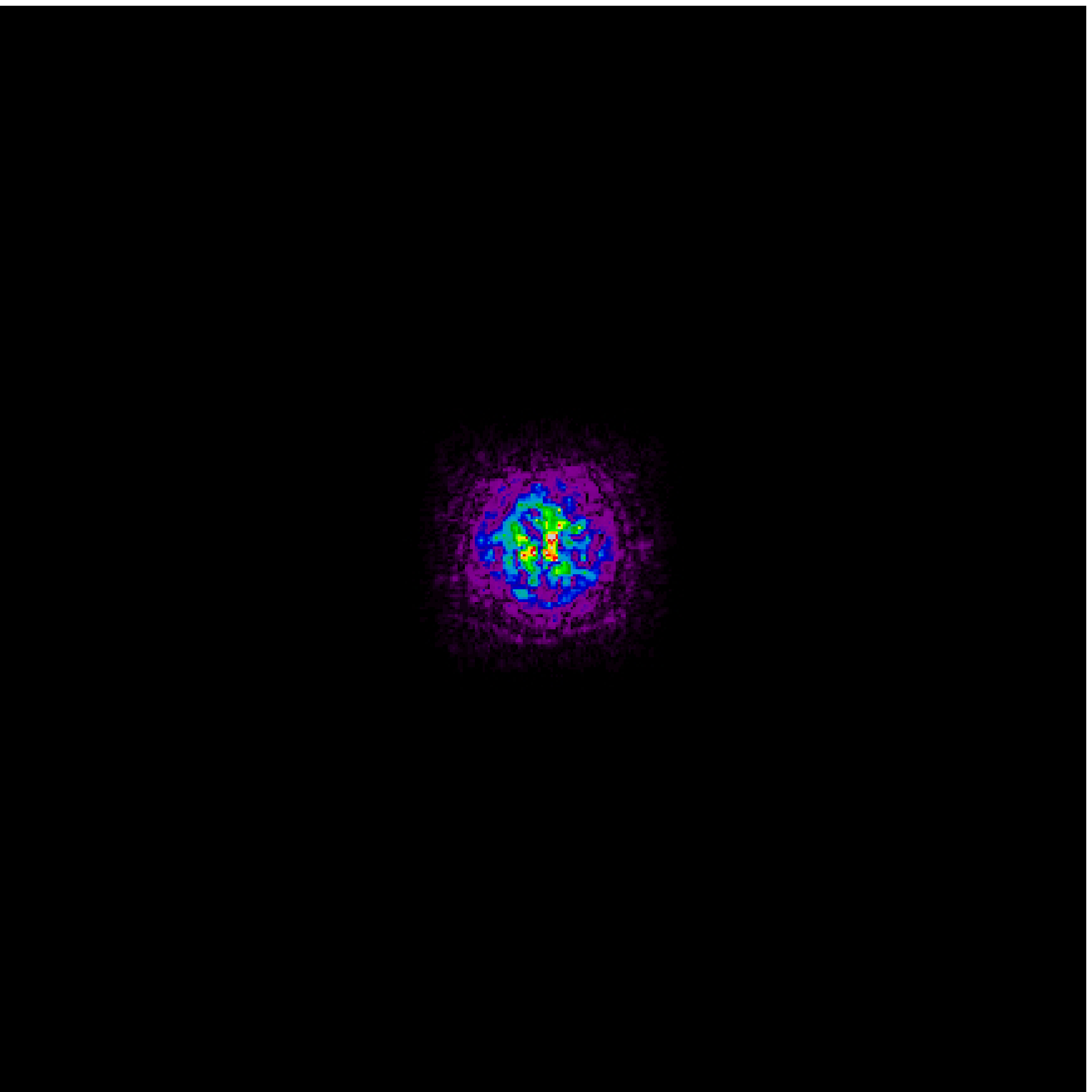}%
\includegraphics[width=0.24\textwidth]{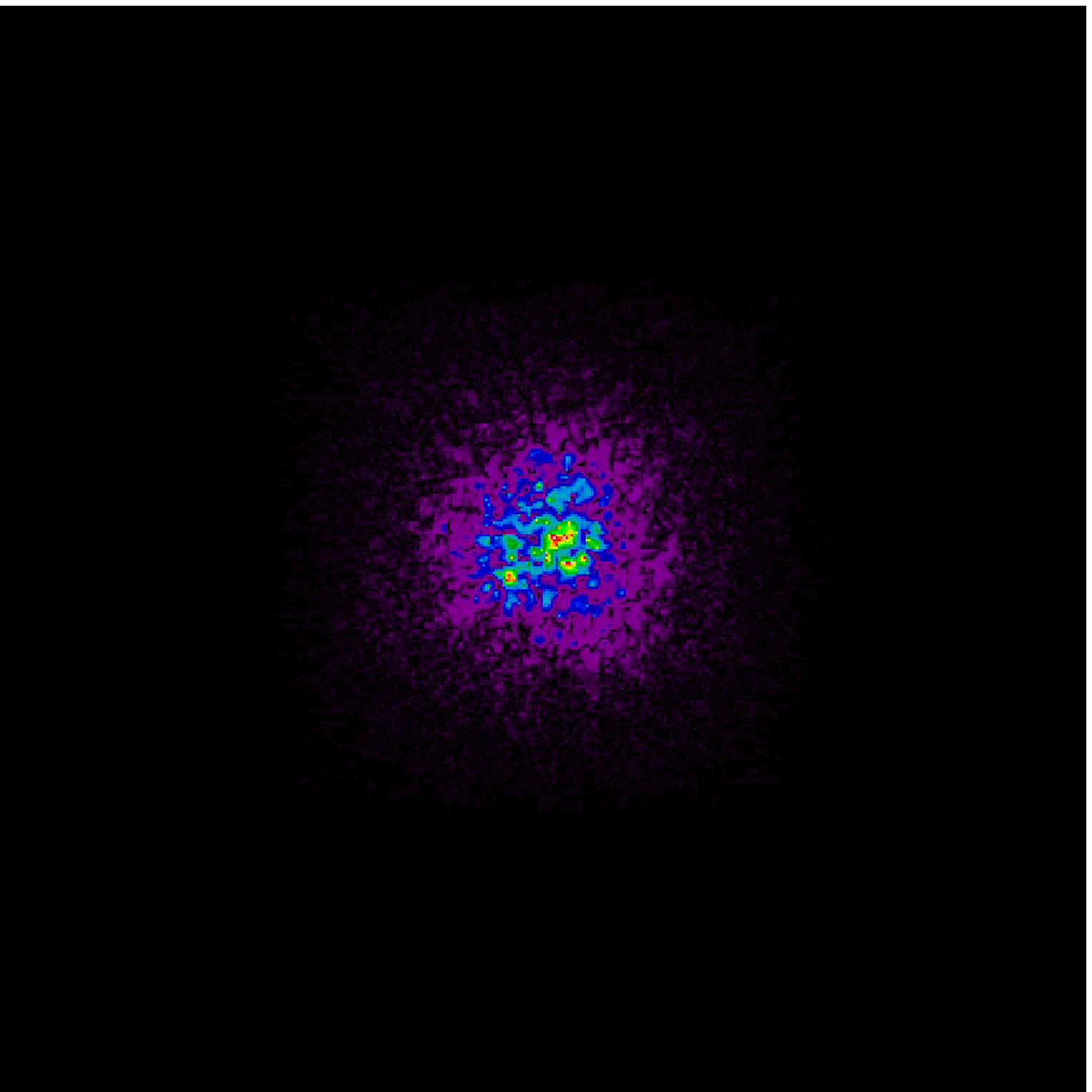}%
\includegraphics[width=0.24\textwidth]{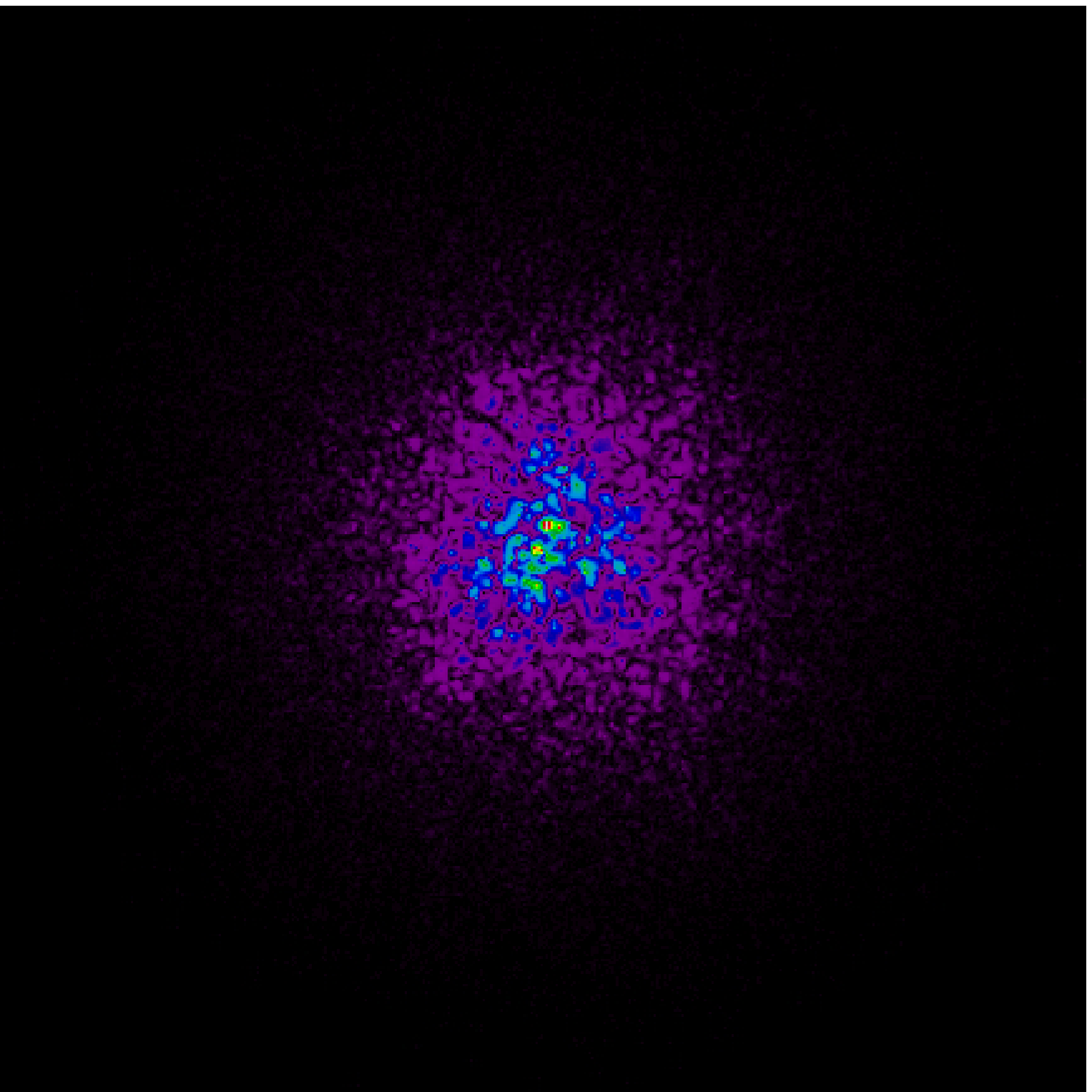}%
\includegraphics[width=0.24\textwidth]{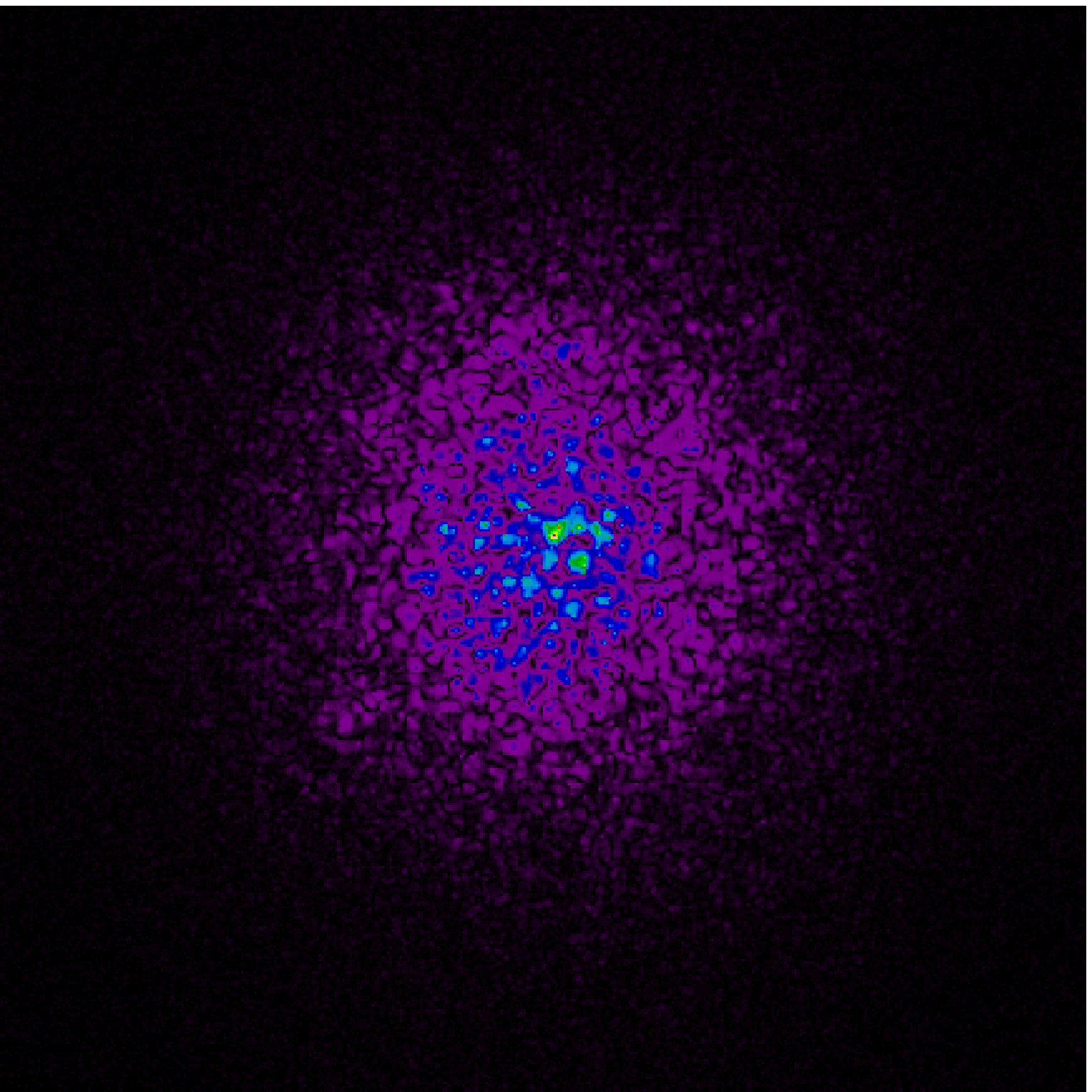}
\caption{Time evolution of the absolute value of the down-spin wavefunction \(\psi_{\!-}\), for \(\tau^{-1}=0.004,\,0.1\). Extended state (top); nearly localized state (bottom). The four panels of each row are for times 20, 60, 160, and 400, respectively.}
\label{f:wave}
\end{figure}

To characterize the geometry of the wavefunction we measure the inverse participation ratio \(P(t)\),\cite{Wegner-1980kx} and the fractal dimension \(D(t)\).\cite{Aoki-1986vn} It is convenient to partition the surface \(L^2\) into \(N_l\) cells of size \(l^2\) and to define the measure,\cite{Chhabra-1989fk}
\begin{equation}\label{e:mk}
m_k(l,t)=\sum_{i=1\in k}^{l^2} \left|\psi_i(t)\right|^2,
	\quad k=1,\dots,N_l\,,
\end{equation}
(\(\sum_k m_k=1)\), from which we write the dynamical \(q\)-inverse participation ratio,
\begin{equation}\label{e:Pq}
P_q(l,t)=\sum_{k=1}^{N_l}m_k(l,t)^q\,,
\end{equation}
and \(q\)-fractal dimension,\cite{Halsey-1986fk}
\begin{equation}\label{e:Dq}
D_q(t)=\lim_{l\rightarrow0} \frac{1}{q-1} 
	\frac{\log P_q(l,t)}{\log l}\,.
\end{equation}
We use in the following the special case with \(q=2\), \(D=D_2\) and \(P=P_2\). The wavefunction fractal dimension \(D=D_2\), determines the asymptotic behavior of the correlation functions, or return probability, \(\sim t^{-D_2/2}\) in the critical state.\cite{Ketzmerick-1997fk} The inverse participation ratio tends to zero when the wavefunction is extended, and remains finite (at long times) when the quantum state is localized; concomitantly, the fractal dimension drops from a value of \(2\) (the entire surface) in the extended regime, to a smaller value in the localized regime. These quantities are well suited for characterizing the Anderson transition as a function of the disorder.\cite{Evers-2008nx}  The relevant parameter to characterize the disorder strength is the scattering time \(\tau\), which is proportional to the exchange energy and the impurities concentration \(c\), \(\tau^{-1}\approx cJ_s^2\).

\section{Results}\label{s:results}

%current
\begin{figure}[t]
\centering
\includegraphics[width=0.48\textwidth]{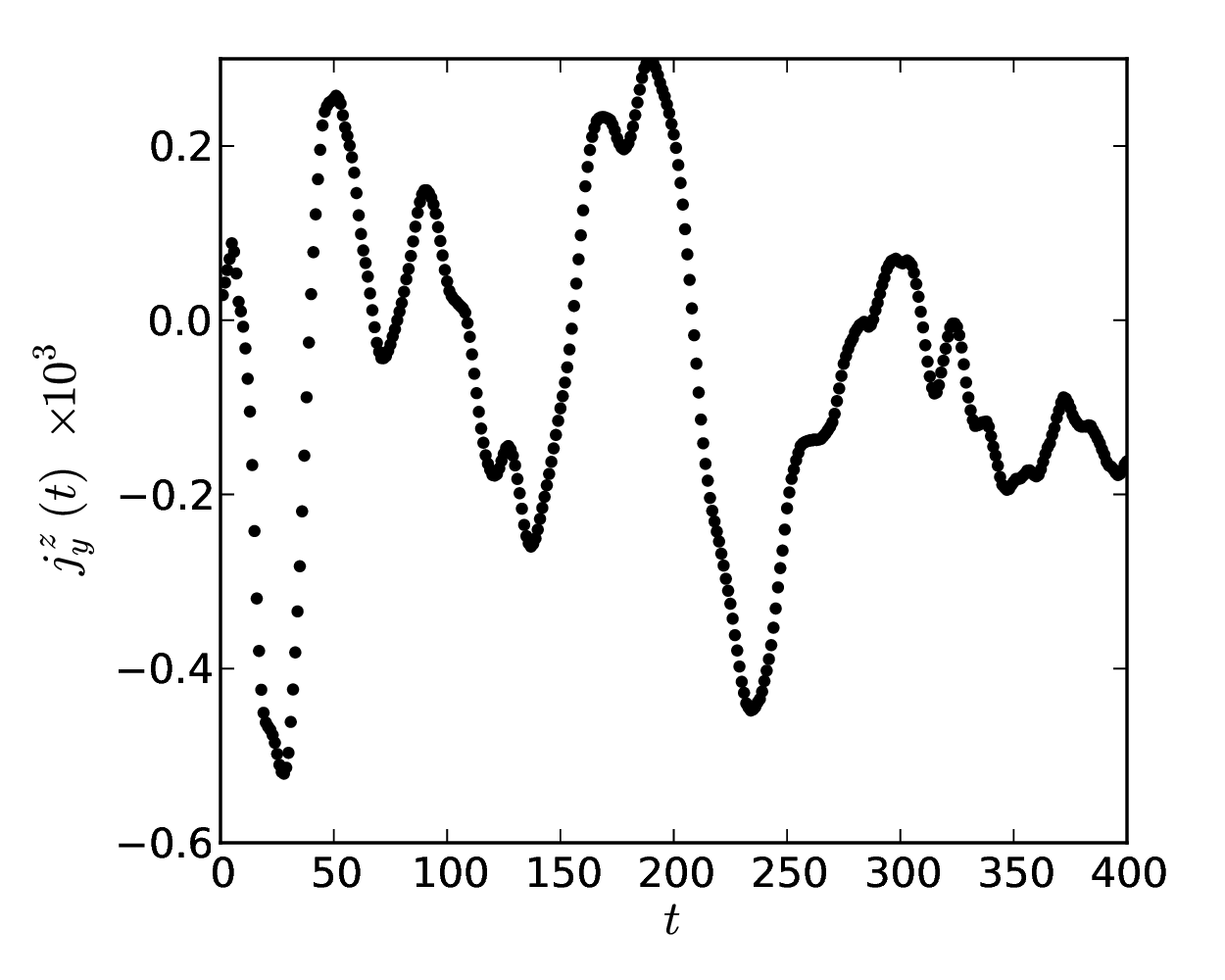}%
\includegraphics[width=0.48\textwidth]{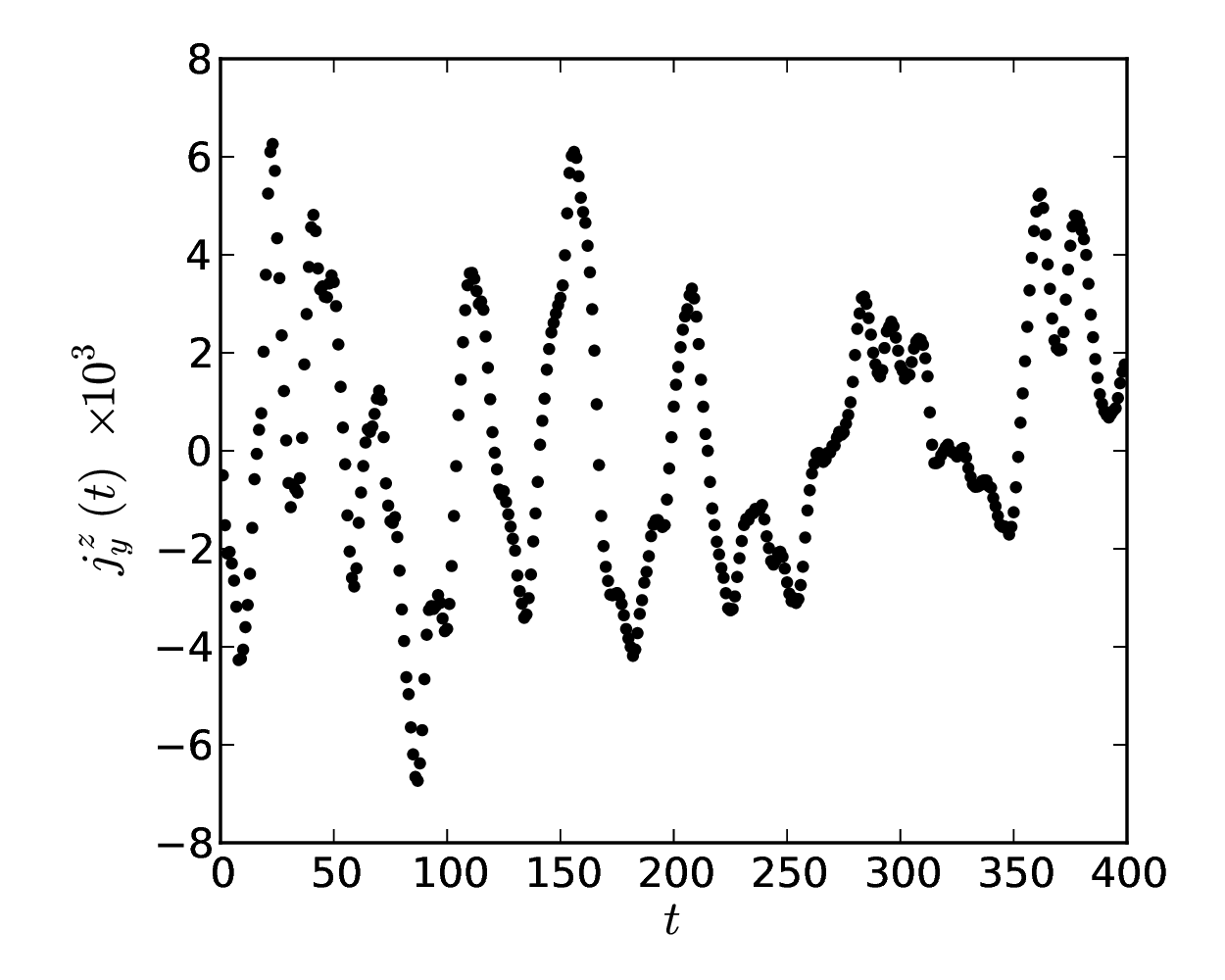}
\caption{Spin current as a function of time, corresponding to \(\tau^{-1}=0.004\) (left) and  \(\tau^{-1}=0.1\) (right), the two cases of Fig.~\protect\ref{f:wave}.}
\label{f:current}
\end{figure}

We analyze the temporal evolution of an initially normalized Gaussian wave packet, concentrated around the origin,
\begin{equation}\label{e:e:psi0}
\psi_{\!+}(\bm x,0) = \frac{1}{(2\pi d^2)^{1/2}}\exp\left(-\frac{|\bm x|^2}{4d^2}\right),
\end{equation}
and  \(\psi_{\!-}(\bm x,0)=0\), where the size is typically \(d=12a\), much smaller than the system size (\(\sim 512 a\)). We vary the value of the disorder strength, essentially by changing the exchange coupling constant. In a clean system, which can be taken as a reference, the width of the wave packet spreads ballistically, \(w\sim t\), its correlation dimension \(D=2\), and its inverse participation ratio tends to zero.

%width
\begin{figure}[t]
\centering
\includegraphics[width=0.48\textwidth]{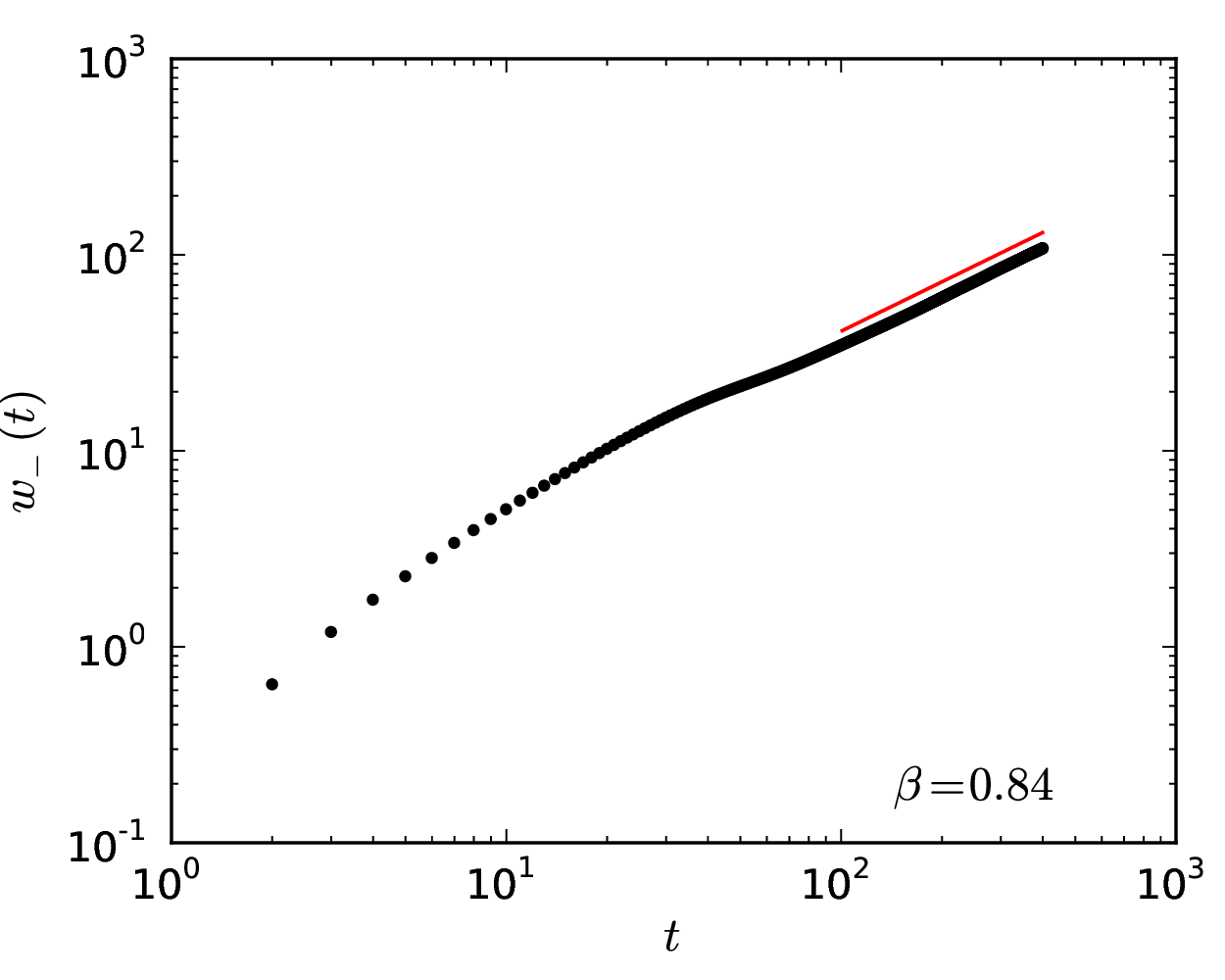}%
\includegraphics[width=0.48\textwidth]{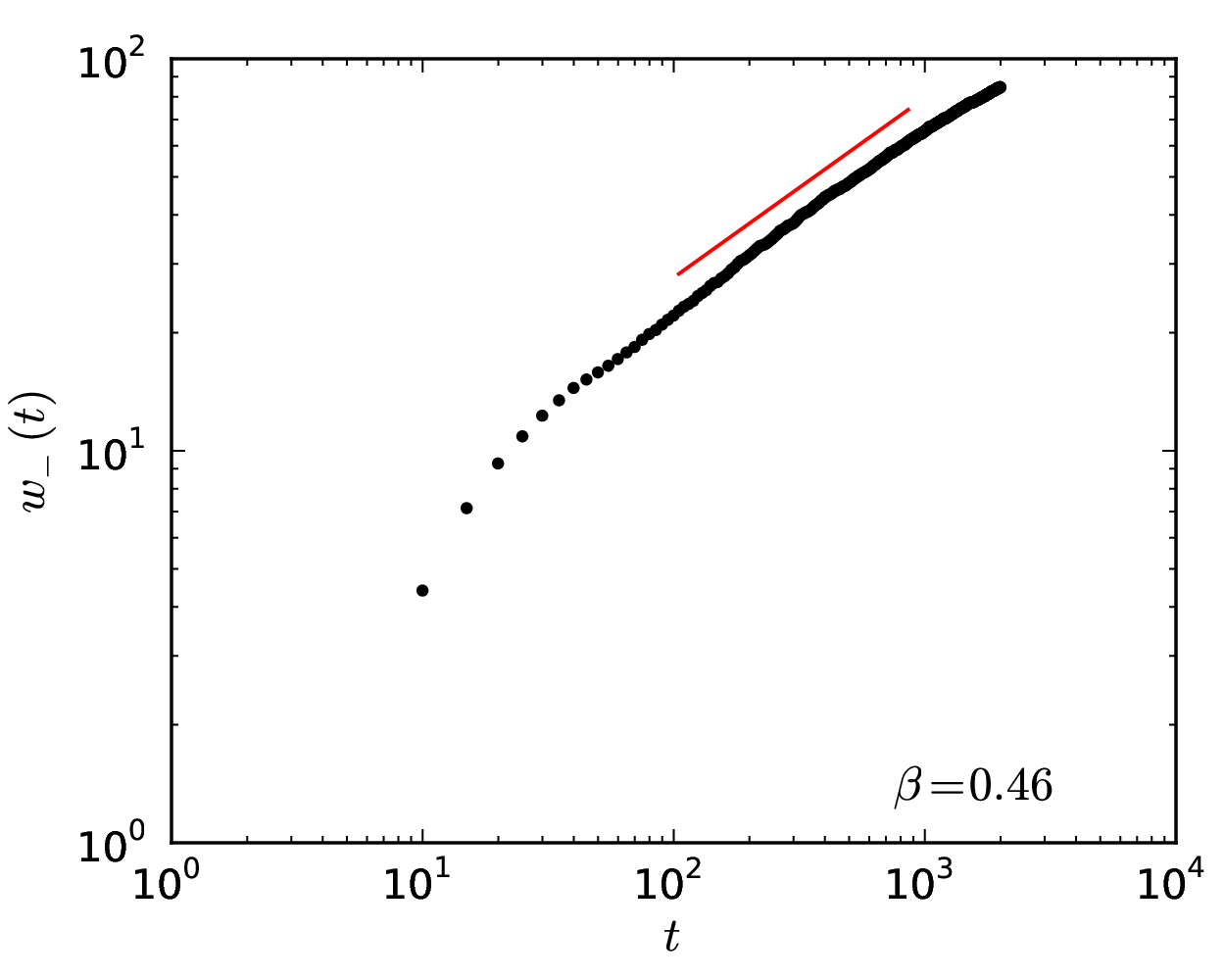}
\caption{Spreading of the wavefunction \(\psi_{\!-}\) for \(\tau^{-1}=0.004\) (left) and  \(\tau^{-1}=0.1\) (right), as in the previous figures. They follow a power law with exponent \(\beta\) (slope of the solid line).}
\label{f:width}
\end{figure}

In Fig.~\ref{f:wave} we compare the time evolution of the Gaussian wave packet for two values of the disorder strength: \(\tau^{-1}=0.1\times (0.2)^2,\,0.1\times (1.0)^2\) (the impurity concentration is \(c=0.1\)). We note, on the one hand, that in the weak disorder case (top) the continuous spreading of the wavefunction is accompanied by a significant amount of interference; on the other hand, for stronger disorder (bottom), the spatial distribution of the wavefunction is more complex, and the spreading slower. Associated with these two regimes, we observe that the spin current (for one realization of disorder) is greatly enhanced in the strong regime, as shown in Fig.~\ref{f:current}. The order of magnitude of the amplification factor is the ratio of the respective scattering strengths. 

This somewhat paradoxical result means that, increasing the disorder and consequently decreasing the electron mobility, the spin current (fluctuations) intensifies. Indeed, the spin current is not directly related to the (ballistic) mobility of the carriers, but to their drift ``velocity'' (spin tilting) in an effective, time dependent, magnetic field created by the cyclotron spin-orbit motion. We remind that in our case, the spin current result from the time dependence of the quantum state, together with the vanishing of the probability current, and hence the charge current (the spin current is invariant under time reversal, at variance with the charge current).

\begin{figure}[t]
\centering
\includegraphics[width=0.48\textwidth]{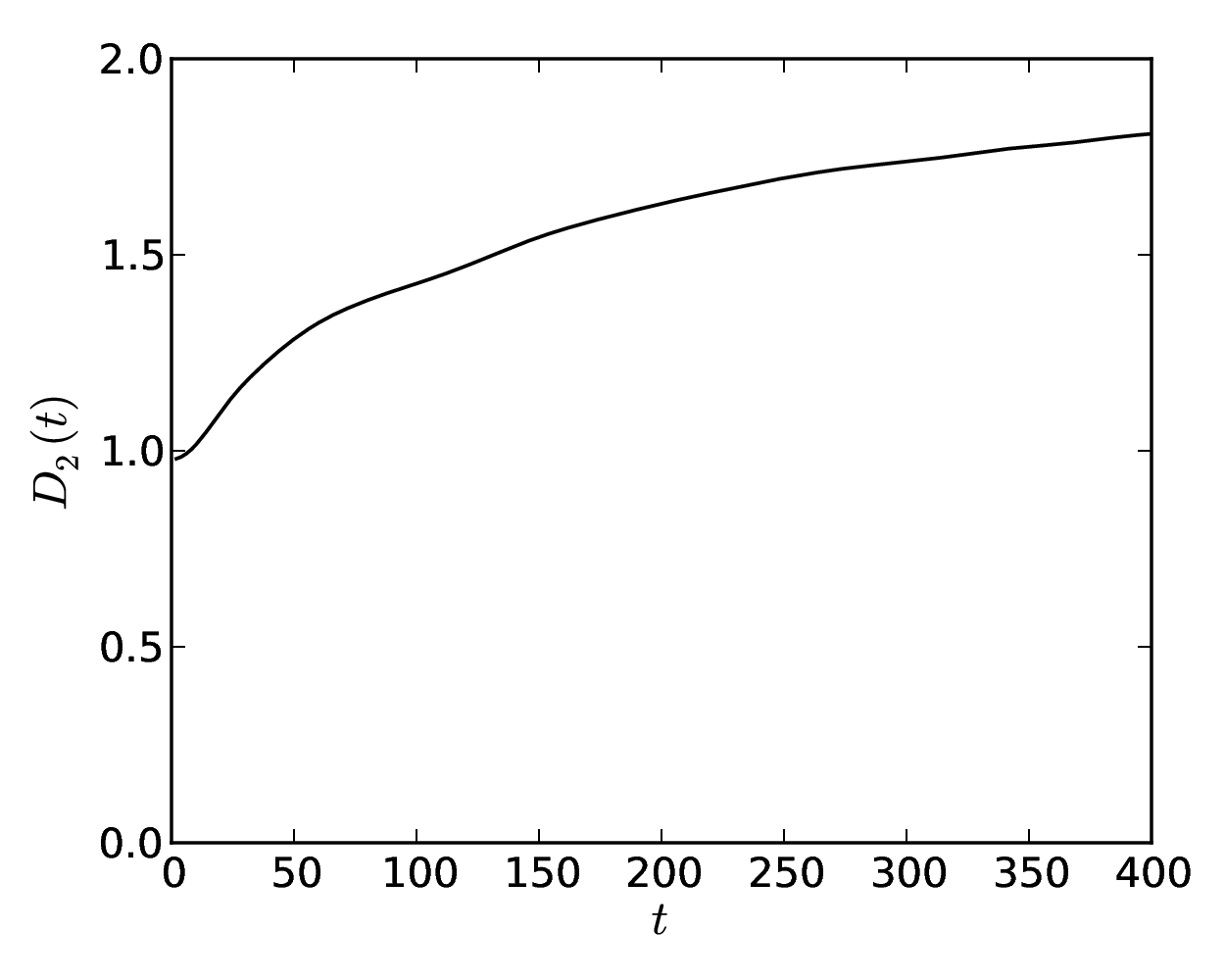}%
\includegraphics[width=0.48\textwidth]{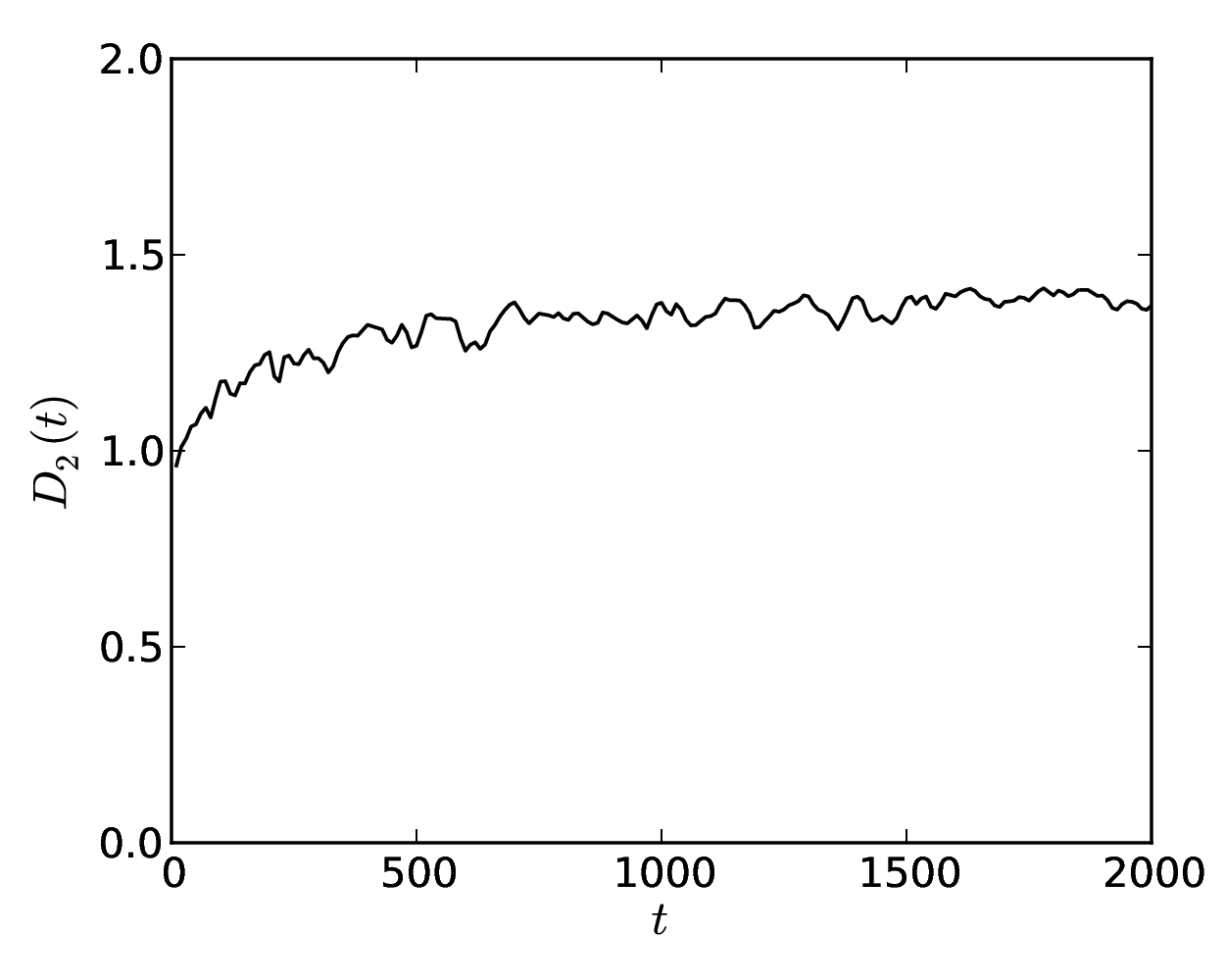}
\caption{Fractal dimension for \(\tau^{-1}=0.004\) (left) and  \(\tau^{-1}=0.1\) (right), as in the previous figures. The strong disorder case is near the localization transition.}
\label{f:fractal}
\end{figure}

\begin{figure}[t]
\centering
\includegraphics[width=0.32\textwidth]{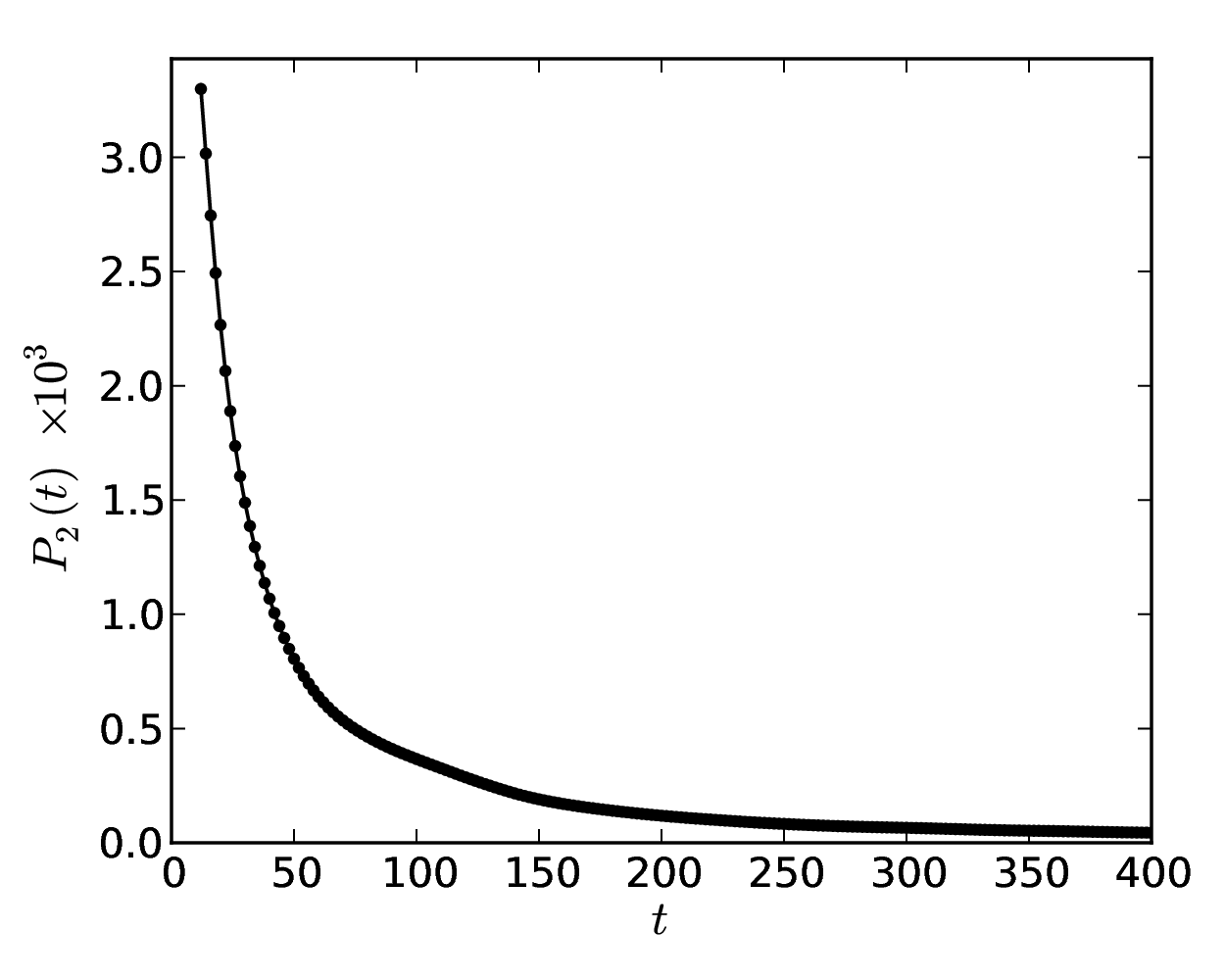}%
\includegraphics[width=0.32\textwidth]{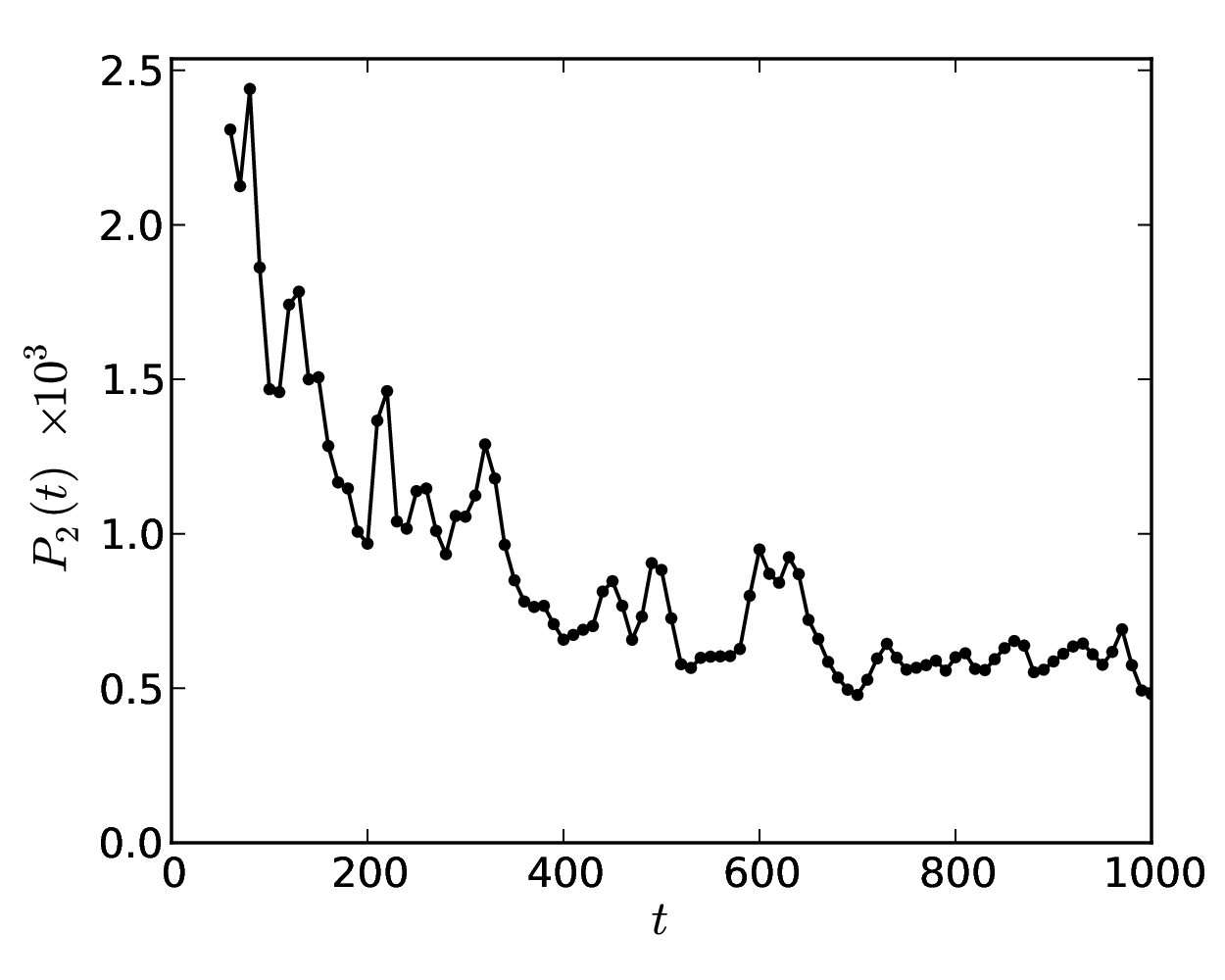}%
\includegraphics[width=0.32\textwidth]{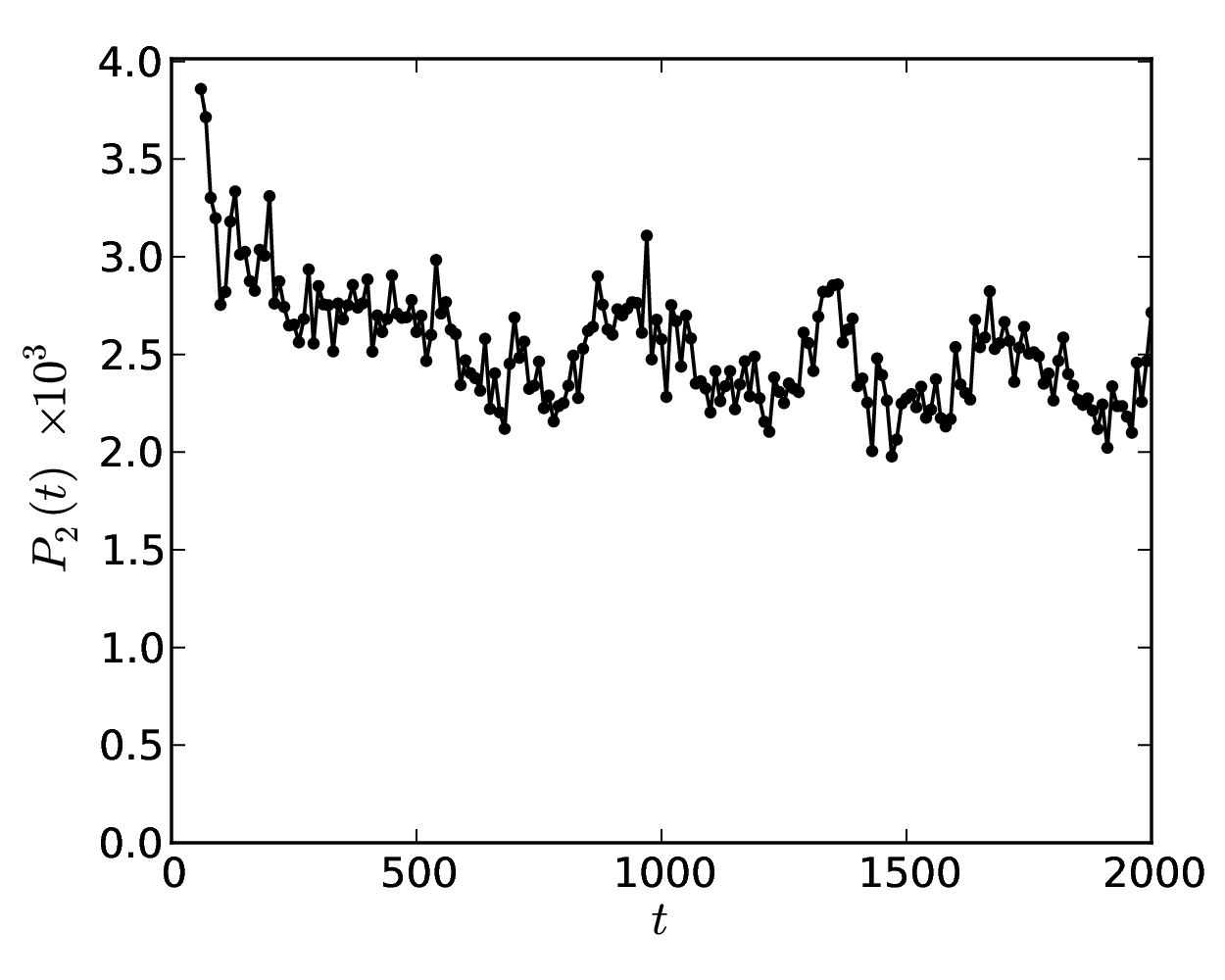}
\caption{Inverse participation ratio, for three scattering times, \(\tau^{-1}=0.004\) (left), \(0.1\) (center), and \(2.5\). Note the different time ranges. The intermediate case is already in the localization regime.}
\label{f:participation}
\end{figure}

These qualitative features are more quantitatively characterized by the measure of the wavefunction width (Fig.~\ref{f:width}), the fractal dimension (Fig.~\ref{f:fractal}), and the inverse participation ratio (Fig.~\ref{f:participation}). We find the the asymptotic behavior of the width is well described by a power law \(w(t)\sim t^\beta\). However, in contrast to the usual ballistic \(\beta=1\) or diffusive behavior \(\beta=0.5\), the exponent explicitly depends on the disorder strength, as seen in Fig.~\ref{f:width}, where \(\beta=0.84\) and \(0.46\). This is confirmed by the measure of the fractal dimension of Fig.~\ref{f:fractal}, that initially increases slowly with time, but remains well below 2 (the value of the extended state, filling the surface) in the strong disorder case. The slower spreading rate can be attributed to a sticking effect of the wavefunction around the impurities, leading simultaneously to a complex spatial distribution (as reflected by the smaller correlation dimension).

%fractal

We may suppose that this later case should be near a localization transition. In order to verify this prediction, we compare the behavior of the inverse participation ratio for three different values of the scattering strength, the two used so far, and a very strong one, \(\tau^{-1}=2.5\), for which we must certainly be in the localized state. We plot \(P(t)\) in Fig.~\ref{f:participation}, for the three cases; we note that already the intermediate case, shows a saturation to a finite value, indicating that the wavefunction is actually localized. The fractal dimension in this case tends to the stationary value of \(D=1.33\) (c.f. Fig.~\ref{f:fractal}). In the very strong disorder case, the fractal dimension saturates at \(D=1.0\). 

Therefore, around the value \(\tau^{-1}\approx 0.1\) a localization transition takes place, characterized by are a drastic reduction in the wave packet spreading rate, an important enhancement of spin current fluctuations, and a correlation dimension that approaches one.

%participation

\section{Conclusion}\label{s:concl}
We demonstrated that the transport properties, diffusion of the wavefunction, spin current, in a two-dimensional semiconductor heterostructure are related to the localization of the quantum state. The interplay of spin-orbit coupling and magnetic disorder leads to atypical behavior of the wavefunction spreading, with a power law that depends on the scattering time in the extended regime. These dependence can be attributed to the importance of interference processes, that partially localize the wavefunction around the impurities. 

For strong disorder, the system undergoes an Anderson transition. The localization of the quantum states is accompanied by the appearance of strong fluctuations in the spin current. In addition, the wavefunction develops a fractal structure with a correlation dimension well below the normal value, corresponding to the system's space dimension, reflecting a highly intermittent probability distribution. These spatial fluctuations allow the existence of large, local spin up and down fluctuations, contributing to spin transport.

It would be worth studying the behavior of the local density of states and its relation with the spin Hall conductivity. This investigation is in progress.

\bibliographystyle{./ws-procs9x6}

\end{document}